 \documentclass[journal,10pt]{IEEEtran}
\usepackage{amsmath}
\usepackage{amsthm}
\usepackage{amsfonts}
\usepackage{amssymb}
\usepackage{makeidx}
\usepackage{cite}
\usepackage{graphicx,bigints}
\usepackage{mathtools}
\usepackage{cases}
\usepackage{xcolor}
\usepackage{hyperref}
\usepackage{bbm}
\usepackage[normalem]{ulem}
\usepackage{braket}
\usepackage{epstopdf}

\newtheorem{theorem}{Theorem}

\newtheorem{lemma}{Lemma}

\newtheorem{example}{Example}

\newtheorem{proposition}{Proposition}
\newtheorem{remark}{Remark}

\DeclareMathOperator*{\argmax}{arg\,max}

\DeclareMathOperator{\cS}{\mathcal{S}}
\DeclareMathOperator{\cF}{\mathcal{F}}

\DeclareMathOperator{\bP}{\mathbf{P}}
\DeclareMathOperator{\ind}{\mathbbm{1}}
\DeclareMathOperator{\bE}{\mathbf{E}}
\DeclareMathOperator{\bZ}{\mathbb{Z}}
\newcommand*\diff{\mathop{}\!\mathrm{d}}

\newcommand*\nnb{\nonumber}

\definecolor{sandy}{HTML}{E6E2AF}
\definecolor{stone}{HTML}{A7A37E}
\definecolor{beach}{HTML}{EFECCA}
\definecolor{ocean}{HTML}{046380}
\definecolor{diver}{HTML}{DA5A2A}

\definecolor{Firenze1}{HTML}{468966}
\definecolor{Firenze2}{HTML}{FFF0A5}
\definecolor{Firenze3}{HTML}{FFB03B}
\definecolor{Firenze4}{HTML}{B64926}
\definecolor{Firenze5}{HTML}{d3687f}

\newcommand{\ea}{\stackrel{(\text{a})}{=}}
\newcommand{\eb}{\stackrel{(\text{b})}{=}}
\newcommand{\ec}{\stackrel{(\text{c})}{=}}
\newcommand{\ed}{\stackrel{(\text{d})}{=}}

\title{A Random Geometric Model of Blockages in Vehicular Networks}
\author{Chang-Sik~Choi and François Baccelli
	\IEEEcompsocitemizethanks{\IEEEcompsocthanksitem 	{Chang-sik Choi is with the School of Electronic and Electrical Engineer, Hongik University, South Korea. François Baccelli is with the University of Texas at Austin, USA and also with Inria Paris, France.  (email: chang-sik.choi@hongik.ac.kr, baccelli@math.utexas.edu, francois.baccelli@inria.fr).} }
	}

\begin{document}
	\maketitle 
\begin{abstract}	
This paper presents a novel spatially consistent approach for modeling line-of-sight (LOS) paths in vehicular networks. We use stochastic geometry to model transmitters, obstacles, and receivers located in three parallel lines, respectively. Their geometric interactions are leveraged to characterize the existence of LOS paths. Specifically, the proposed approach focuses on the role of obstacles in blocking one or more LOS paths, which has been overlooked in most statistical models for blockage. Under the proposed framework, we derive the probability that a typical vehicle is in LOS with respect to transmitters with received signal-to-noise ratios greater than a threshold. The proposed framework and LOS coverage analysis are instrumental to the analysis of LOS-critical applications such as positioning or mmWave communications in vehicular networks. 
\end{abstract}


\section{Introduction}
\subsection{Motivation and Background}
Line-Of-Sight (LOS) paths play a key role in various communication systems. For instance, in positioning systems, users or vehicles can estimate their relative distances and positions based on the time-of-arrivals (ToAs) and time-of-departures (ToDs) of LOS signals from various transmitters \cite{1611097,1618619,5208736,fischer2014observed}. In mmWave communication systems, beams are directional, which leads to significant performance fluctuations. Nevertheless, users with access to LOS path signals can achieve a high data rate \cite{6736750,7109864}. In such wireless systems, obstacles may obstruct the direct LOS paths and this results in ranging and positioning error, and possibly unstable communications \cite{fischer2014observed, 10.1145/3366423.3380169}. Consequently, it is essential to accurately understand the LOS and blockage probabilities in such systems.

Various papers, including \cite{6290250,38901,7546922,8666148,8610083}, studied the statistical behavior of LOS and blockage. The independent blockage models in \cite{6290250,38901} have been widely used for their simplicity, as the LOS profile is independently created for each transmitter-receiver pair. In other words, nearby or even co-located devices are forced to have independently created LOS profiles and this results in a LOS and blockage model that is not based on the geometric interactions between obstacles and transmitter-receiver pairs. Nevertheless, from first principles, the LOS property is merely a manifestation of spatial interactions between network components. For instance, if the direct path between a transmitter and a receiver is not blocked by an obstacle, the receiver is in LOS. In a similar way, if two receivers are very close, they are likely to experience the same obstacles. It is essential to develop a model that simultaneously captures LOS and blockage. For instance, in \cite{7563397}, such a spatially consistent channel model was considered for mmWave in-building networks. In the same vein, our aim is to develop a spatially consistent approach to evaluate LOS probabilities by incorporating the geometric interactions between obstacles, receiver vehicles, transmitters in this context. 

\section{System Model}\label{S:2}

\begin{figure}
	\centering
	\includegraphics[width=1\linewidth]{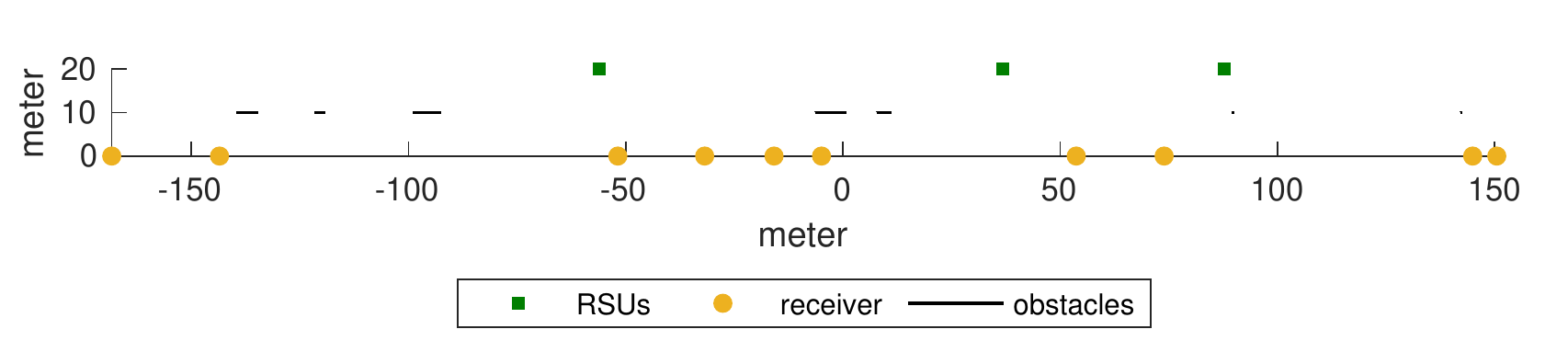}
	\caption{Here, we use $ \lambda_r = 10/\text{km} $, $ \lambda_b = 20/\text{km}, $ $ \lambda_v=30/\text{km} $, and $ 1/\mu=2.5 $ meters. }
	\label{fig:fig00}
\end{figure}
\begin{figure}
	\centering
	\includegraphics[width=1\linewidth]{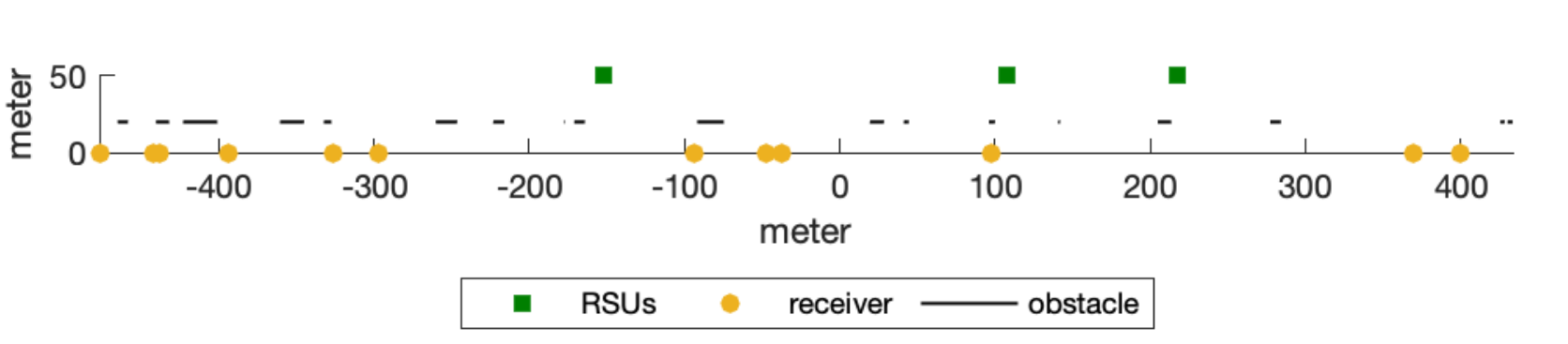}
	\caption{Here, we use $ \lambda_r = 5/\text{km} $, $ \lambda_b = 20/\text{km}, $ $ \lambda_v=10/\text{km} $, and $ 1/\mu=5 $ meters. }
	\label{fig:fig01}
\end{figure}

\begin{figure}
	\centering
	\includegraphics[width=.9\linewidth]{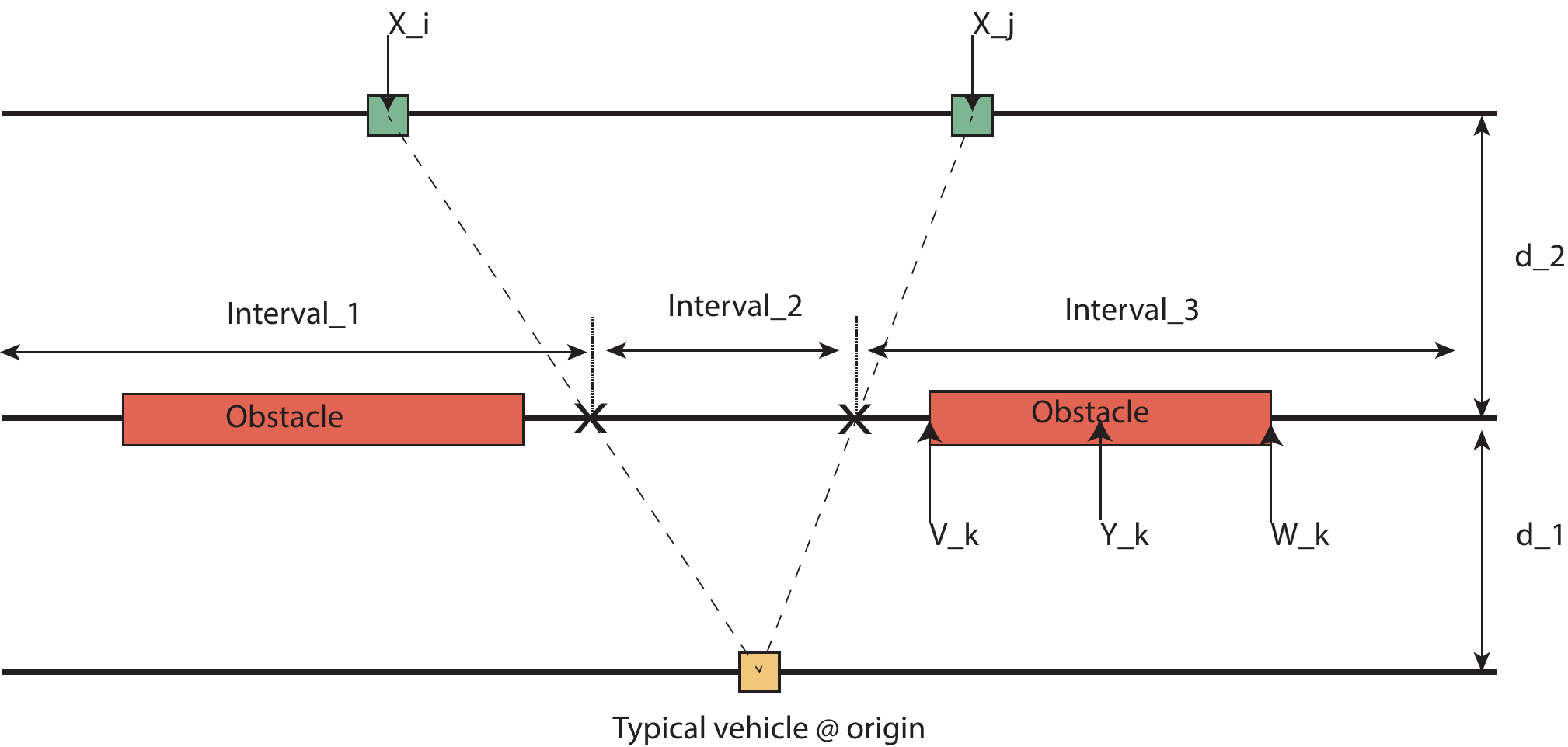}
	\caption{Transmitters at $ (x_i,d_1+d_2) $ and $ (x_j,d_1+d_2) $ and their projections onto $ y=d_1+d_2 $. For the analysis of blockage from obstacles, we assume that obstacles are of width zero.}
	\label{fig:picture2}
\end{figure}

\begin{figure}
	\centering
	\includegraphics[width=.9\linewidth]{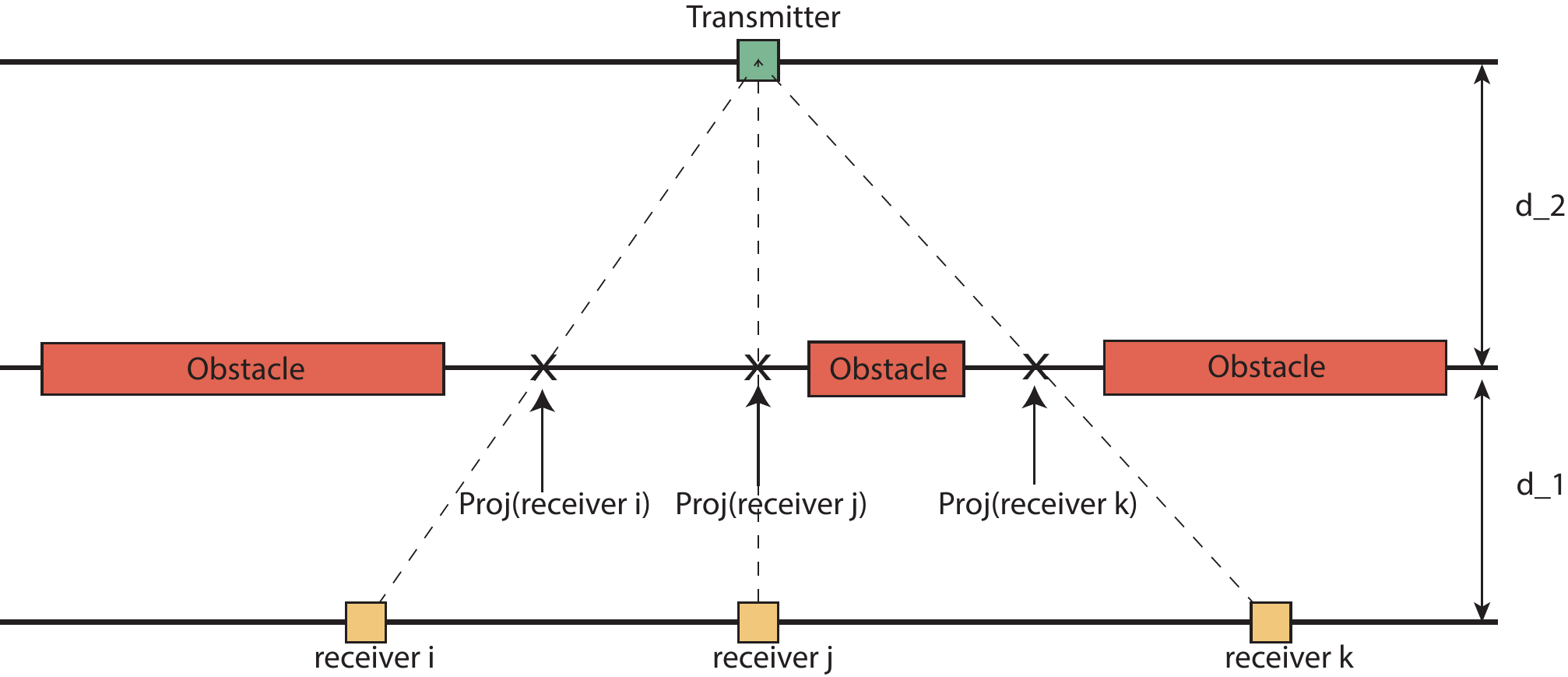}
	\caption{Receivers at line $ y=0 $ and their projections onto $ y=d_1 $. }
	\label{fig:picture4}
\end{figure}

\subsection{Spatial Model}

We first model the random locations of vehicle receivers as a Poisson point process $ \Phi_v  = \{Z_i\}_{i\in\bZ}$ with intensity $ \lambda_v $ on the $ x $-axis. In this model, the $ x $-axis corresponds to the road on which the receiving vehicles move. Specifically, each vehicle receiver decides its moving direction---positive or negative $ x $-direction---according to a Bernoulli distribution ($ p=0.5 $). Their speeds are assumed to be constant and equal to $ v\geq 0 $. 

\par The locations of transmitters---base stations or roadside units---are modeled as an independent Poisson point process $ \Phi_r = \{X_i\}_{i\in\bZ} $ with intensity $ \lambda_t $ on the line $ y=d_1+d_2, $ where $ d_1, d_2 \geq 1 $. Transmitters are assumed to be static.

%

\par
\begin{figure}
	\centering
	\includegraphics[width=1\linewidth]{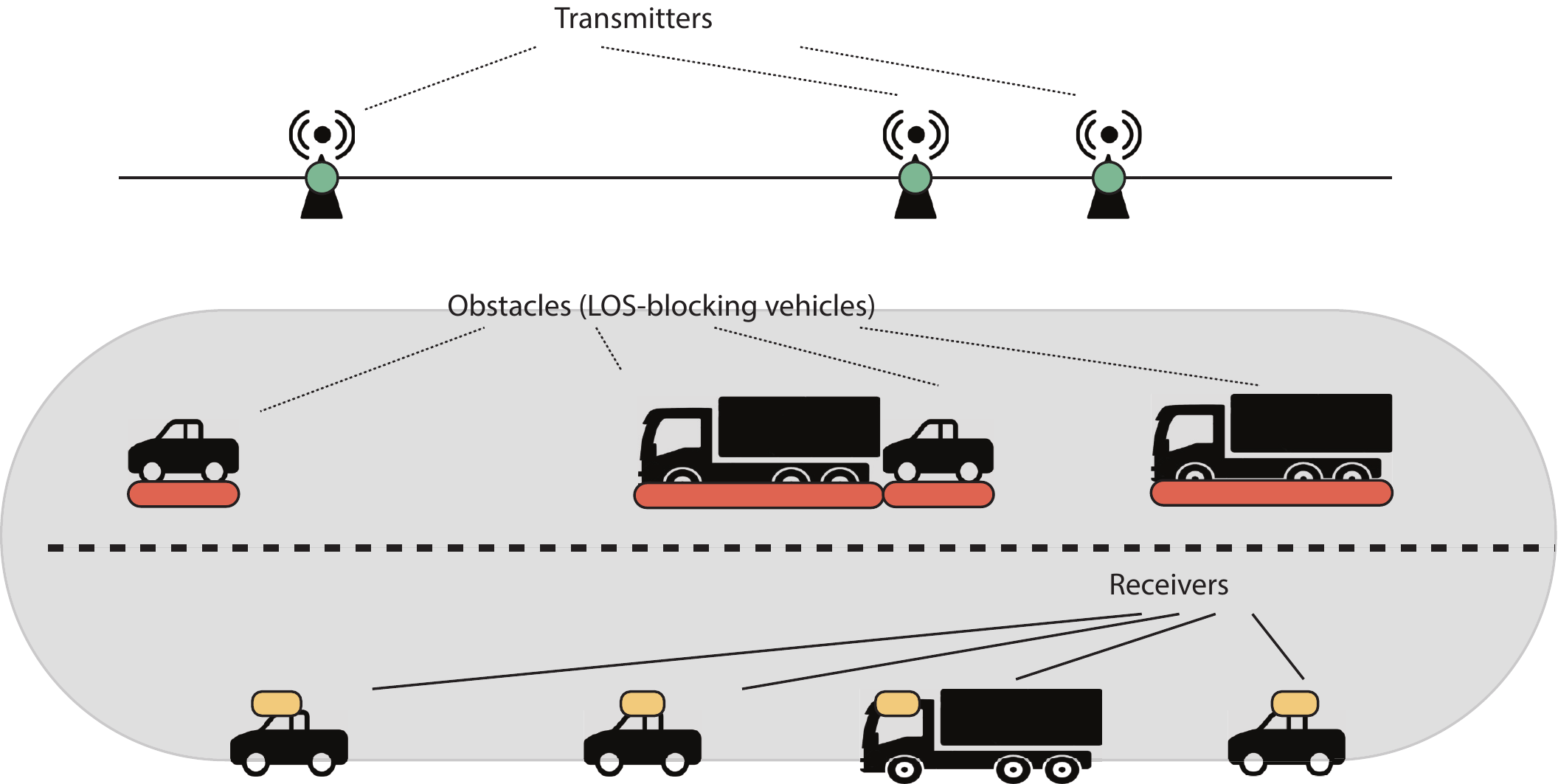}
	\caption{Illustration of the vehicular networks where obstacles block the direct LOS paths from transmitters to vehicle receivers.}
	\label{fig:vehicleslosblocking}
\end{figure}

 Potential LOS-blocking obstacles---such as vehicles on different lanes---are modeled as one-dimensional segments parallel to the $x $-axis. Specifically, we use the fact that in vehicular networks, LOS-blockages are caused by obstacles of various lengths. For analytical tractability, we model their {centers} as an independent Poisson point process $ \Phi_b  = \{Y_j\}_{j\in\bZ}$ of intensity $ \lambda_b $ on the line $ y=d_1 $. Then, based on these centers, each obstacle $ \cF_j $ is modeled as a one-dimensional i.i.d. length random segment parallel to the $ x$-axis.  The left and right points of obstacle $ \cF_j $ are denoted by $ V_j $ and $ W_j, $ respectively. See Figs.  \ref{fig:fig00} to \ref{fig:vehicleslosblocking}. To capture the blocking obstacles' variable sizes and lengths, we propose to model the half-lengths of each obstacle---namely $ \tilde{V}_j=\|Y_j-V_j\|$ and $ \tilde{W}_j=\|Y_j-W_j\| $---as i.i.d. exponential random variables with mean $ 1/\mu $. Therefore, the full length of each obstacle follows an Erlang-$ 2  $ distribution with mean $ 2/\mu $.  Note that this modeling of obstacles' length is proposed for analytical tractability. Similar to the motion of vehicles, obstacles are assumed to move at constant speed $ v_o\geq 0, $ where their moving directions on the line $ y=d_1 $ are independently and uniformly chosen at time $ 0. $ As a result, at any given time, the set of obstacles is  a Boolean model with respect to the obstacle point process $ \Phi_b $ \cite{baccelli2010stochastic}. 

\begin{figure}
	\centering
	\includegraphics[width=0.7\linewidth]{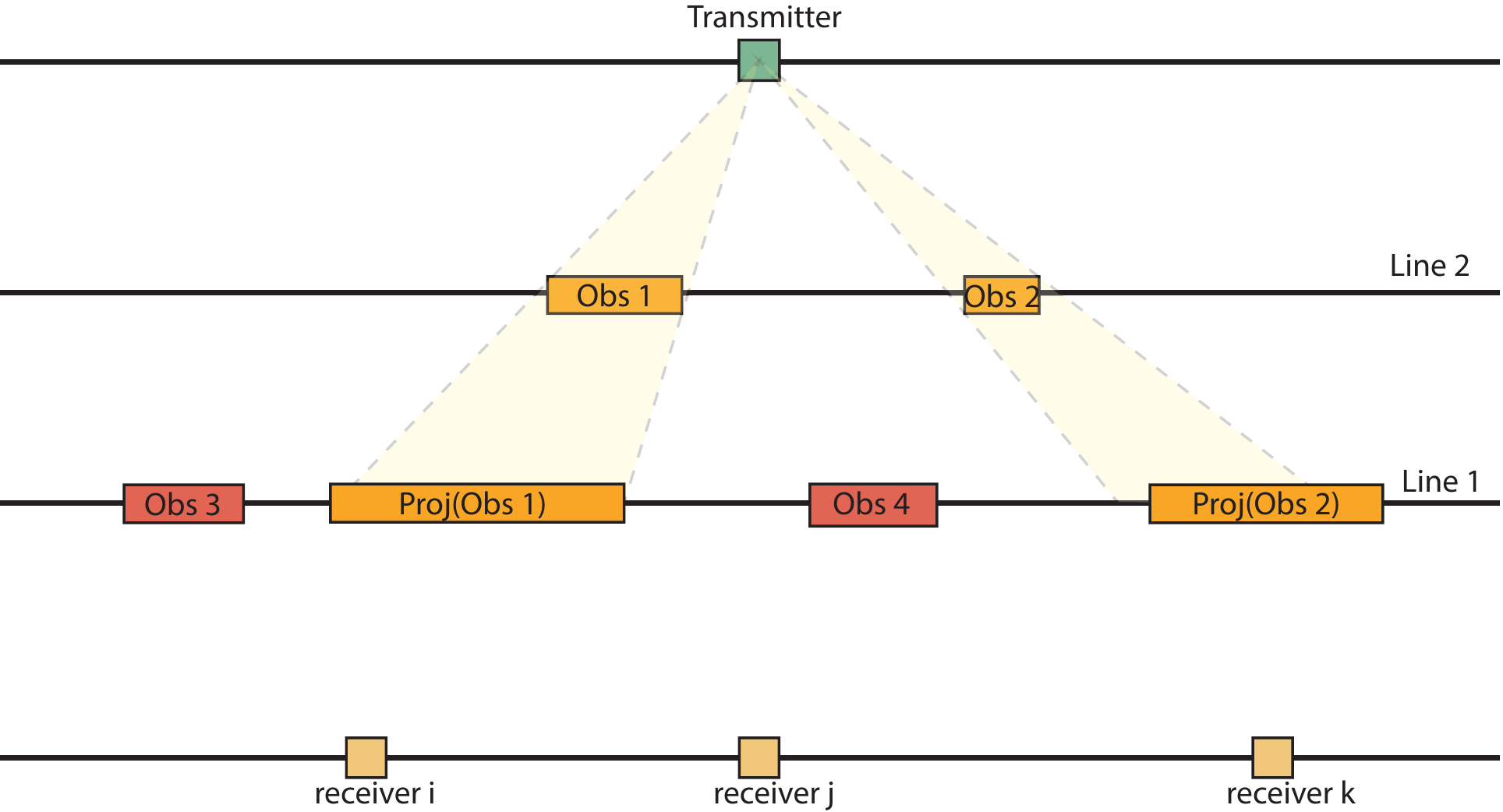}
	\caption{Illustration of superposition of Boolean models on different lines. Here, obstacles $ 1 $ and $ 2 $ are projected to the line $ 1  $. The projections of a Boolean model onto line $ 1 $ form a Boolean model. }
	\label{fig:picture5}
\end{figure}

\begin{remark}
In this paper, we assume that all obstacles are located on a single line. Nevertheless, obstacles located on multiple parallel lines can be analyzed within the same framework. Specifically, since we focus on direct paths between transmitters and receivers, the obstacles on different lanes can be projected onto a single line to characterize the blockage of direct LOS paths. For instance, when obstacle centers on different lines are modeled as independent Poisson point processes and obstacles' sizes are modeled as i.i.d. exponential random variables, the projections of obstacles can also be seen as a Boolean model created by the superposition of independent Boolean models \cite{baccelli2010stochastic,baccelli2010stochasticvol2,chiu2013stochastic}.  See Fig. \ref{fig:picture5} for the projection of the Boolean models. 
\end{remark}

\subsection{LOS and k-LOS Coverage Probability}
A vehicle is LOS with respect to (w.r.t.) a transmitter $ X $ if the direct path from $ X $ to it is not obstructed by any obstacle. In a similar way, a vehicle is in non-line-of-sight (NLOS) if the direct path is obstructed.

\par In addition to the investigation of the existence of geometric direct LOS paths, we use the attenuation of signals to further characterize LOS blockage. First, by averaging out small-scale fading, we assume that the averaged received signal-to-noise-ratio (SNR) of a LOS receiver at a distance $ d $ is given by $ \frac{p d^{-\alpha_{\text{LOS}}}}{\sigma} $
	where $ p $ the transmit power, $ \alpha_{\text{LOS}} $ the path loss exponent for a LOS channel,  
	 and $ \sigma $ the noise power. 
	 Then, we assume that a transmitter is \emph{detectable} by the typical receiver if $ \frac{p d^{-\alpha_{\text{LOS}}}}{\sigma} $ is greater than a given threshold, namely if it would be detected when LOS. 
	
	\par We define a receiver as being in full LOS coverage if it is in LOS w.r.t. all of its \emph{detectable} transmitters. Similarly, a receiver is in $ k$-LOS coverage if it is LOS w.r.t. at least $ k $ of its  \emph{detectable} transmitters. In this paper, we consider a typical receiver at the origin and derive the probability that it is in full LOS coverage and is in $ k $-LOS coverage, respectively. Since receivers are assumed to be distributed as a Poisson point process, the typical LOS coverage probability seen from the typical receiver corresponds to the LOS coverage spatially averaged over all receivers in the $ x $-axis, as we shall see. 
\begin{remark}
The $ k $-LOS coverage probability is proposed to address various use cases in LOS-critical wireless systems. For instance, in a mmWave cellular network, the significant penetration loss makes it necessary for users to have at least one direct LOS base station \cite{niu2015survey,10.1145/3366423.3380169}. Similarly, for triangulation and positioning systems which leverage  ToAs and ToDs, users are able to accurately localize themselves when two or more LOS paths are available to them  \cite{1611097,1618619,5208736,fischer2014observed}. Full coverage is also useful for advanced positioning systems where multiple time-series of ToAs and ToDs are fused to differentiate and isolate LOS paths out of NLOS paths  \cite{4802193,5555901}. 
\end{remark}

\section{Main Results}
\subsection{LOS Probability}
We first evaluate the probability of having direct LOS paths. 
\begin{lemma}
The transmitter point process $ \Phi_r $, the set of obstacles $ \cF $ and the receiver point process $ \Phi_v $ are jointly stationary, i.e., invariant w.r.t. translation on the $ x $-axis. 
\end{lemma}
\begin{IEEEproof}
The set $ \cF $---collection of obstacles---is stationary because the centers of the obstacles are stationary and the lengths of all obstacles are modeled as an i.i.d. sequence. Therefore, $ \cF $ is a stationary Boolean model\cite{baccelli2010stochastic} .   
\end{IEEEproof}

\begin{proposition}\label{T:1}
	The probability that a typical vehicle receiver at the origin is LOS w.r.t. a transmitter at $ (x,d_1+d_2)$ is $ e^{-2\lambda_b/\mu} $. Correspondingly, the NLOS probability is $ 1-e^{-2\lambda_b/\mu}.$
\end{proposition}
\begin{IEEEproof} Let $ \text{LOS}(x) $ denote the event that the typical vehicle at the origin is LOS w.r.t. the  transmitter at $ (x,d_1+d_2) $. By using an indicator function, we write $ \bP(\text{LOS}(x)) = 1- \bP(\text{NLOS}(x))= 1-\bE[\ind_{\text{NLOS}(x)}]. $ Furthermore, the typical vehicle is NLOS w.r.t. the transmitter  if and only if the direct path is blocked by some obstacle $ \cF_j\in\cF. $ Let $ \ind_{\text{LOS}(x;Y_j) } $ be an indicator function that takes a value of one if the path from the transmitter to the typical vehicle is not blocked by obstacle $ \cF_j $ centered at $ Y_j. $ Then, we have 
	\begin{align}
		\bE\left[\ind_{\text{NLOS}(x)}\right] 
		&\ea\bE\left[\bE\left[1-\ind_{\text{LOS}(x;y_j \forall y_j\in\Phi_b) }|\Phi_b\right]\right]\nnb\\
		&\eb\bE\left[1-\prod_{y_j\in\Phi_b}\bE\left[\left.\ind_{\text{LOS}(x;y_j) }\right|\Phi_b\right]\right]\nnb\\
		&=1-\bE\left[\prod_{y_j\in\Phi_b}\bE\left[\ind_{\text{LOS}(x;y_j) }|\Phi_b\right]\right]\nnb.
	\end{align} 
We obtain (a) by conditioning on $ \Phi_b $ and (b) from the independence of the Poisson point process, respectively. Conditional on the center of obstacles $ \Phi_b $, the left and right end points of each obstructor are denoted by $ \{V_j\}, $ and $ \{W_j\} $, respectively. See Fig. \ref{fig:picture2}. The distances from $ Y_j $ to $ V_j $ and to $ W_j $---denoted by $ \tilde{V}_j $ and $ \tilde{W}_j $---are given by i.i.d. exponentials with mean $ 1/\mu. $ Then, for all $ j $ such that $ y_j < \frac{d_1 x }{d_1 + d_2}, $ the direct path is not blocked by the obstacle at $ y_j $ iff its right end point $ W_j $ is less than $ \frac{d_1 x }{d_1 + d_2} $, i.e., $ \tilde{W}_j<\frac{d_1 x }{d_1 + d_2}-y_j. $ Similarly,  $ \forall j $ such that $ y_j > \frac{d_1 x }{d_1 + d_2}, $ the direct path is not blocked by the obstacle at $ y_j $ iff its left end point $ V_j $ is greater than $ \frac{d_1 x }{d_1 + d_2} $, i.e.,  $ \tilde{V}_j>y_j - \frac{d_1 x }{d_1 + d_2}. $ Since $ \tilde{V}, \tilde{W} $ are exponential with mean $ {1}/{\mu}$, we have  
	\begin{align}
	    &\bE[\ind_{\text{LOS}(x;y_j)}|\Phi_b]\nnb\\
    &=\begin{cases}
    	1-\exp\left(-\mu\left(\frac{d_1x}{d_1+d_2}-y_j\right)\right) &\forall j \text{ s.t. } y_j < \frac{d_1 x}{d_1+d_2},\\
    	1-\exp\left(-\mu\left(y_j - \frac{d_1x}{d_1+d_2}\right)\right) &\forall j \text{ s.t. } y_j > \frac{d_1 x}{d_1+d_2}.\nnb
    \end{cases}
    	\end{align}
    Then, the NLOS probability is given by
    \begin{align}
&\bE\left[\ind_{\text{NLOS}(x)}\right] \nnb\\
& = 1- \bE\left[\prod_{y_j\in\Phi_b}^{y_j<\frac{d_1x}{d_1+d_2}}\left(1-e^{-\mu\left(d_1x/(d_1+d_2)-y_j\right)}\right)\right.\nnb\\
&\hspace{15.5mm}
\left.\prod_{y_j\in\Phi_b}^{y_j>\frac{d_1x}{d_1+d_2}}\left(1-e^{-\mu\left(y_j-d_1x/(d_1+d_2)\right)}\right)\right]\nnb\\
&\ec1-\exp\left(-\lambda_b\int_{\frac{d_1x}{d_1+d_2}}^{\infty}e^{-\mu(y -d_1x/(d_1+d_2))}\diff y\right)\nnb\\
&\hspace{10mm}\times 
\exp\left(
-\lambda_b\int_{-\infty}^{\frac{d_1x}{d_1+d_2}}e^{-\mu(d_1x/(d_1+d_2)-y)}\diff y\right)\nnb\\
&\ed1-\exp\left(-2\lambda_b\int_{0}^{\infty}e^{-\mu u}\diff u\right)\nnb\\
&=1-e^{-2\lambda_b /\mu}.\nnb
\end{align}
To obtain (c), we use the probability generating functional of the obstacle point process of intensity $ \lambda_b $ on line $ y=d_1 $ \cite{baccelli2010stochastic}. We have (d) from a change of variables.  
\end{IEEEproof}
\begin{remark}
The LOS analysis can also be understood by the use of projections of transmitters and receivers onto the line  $ y=d_1 $. For instance, a transmitter on the line $ y=d_1+d_2 $ is projected onto line $ y=d_1 $ where the projection is defined as the intersection of the path from the transmitter to the typical receiver at the origin and the line $ y=d_1. $ See Fig. \ref{fig:picture2}. Then, the probability that the typical receiver is LOS w.r.t. a transmitter coincides with the probability that its projection is not contained in the obstacle Boolean model. In a similar way, receivers on the line $ y=0 $ are projected onto the line $ y=d_1 $ where the projections are defined by the intersections of the paths from receivers to a tagged transmitter at $ (x,d_1+d_2)  $ and the line $ y=d_1. $ See Fig.  \ref{fig:picture4}. 
\end{remark}
 Based on the stationarity of the vehicle receiver point process, the derived LOS probability of the typical receiver coincides with the statistical average of the LOS profiles of all vehicle receivers $ \Phi_v $ over the line $ y=d_1 $ (ergodic theorem) \cite{baccelli2010stochastic}. This follows from the fact that  the projections of receiver vehicles form an independent stationary Poisson point process of intensity $ {\lambda_v}\frac{d_1+d_2}{d_2}$  on line $ y=d_1 $. Therefore, among all vehicle receivers, a fraction of $ e^{-2\lambda_b/\mu} $ are in LOS w.r.t. any typical transmitter, and $ 1- e^{-2\lambda_b/\mu} $ are in NLOS, respectively. 
 \par In addition, under the proposed dynamics of vehicles, the joint probability distribution for the receiver and obstacle point processes at time $ 0 $ and time $ t $ are the same \cite{8419219,9354063}. As a result, the obtained LOS probability w.r.t. transmitter $ (x, d_1+d_2) $ also corresponds to the time-average of the LOS profile of a single receiver as it moves along the $ x $-axis. 
%

\begin{remark}\label{R:3}
	Proposition \ref{T:1}  gives the probability that the obstacle Boolean model contains a single point, where the point is the projection of the transmitter onto $ y=d_1 $. Since $ \cF $ is stationary, the probability for the Boolean model to contain a single point is the same as the volume fraction of the Boolean model \cite{chiu2013stochastic}. 
\end{remark}
The derived LOS probability is a function of the density of obstacles and the average size of the obstacles, and it is not a function of the location of the transmitter. We observe that with a single transmitter, the LOS probability is decreasing exponentially with the density of obstacles and their average lengths. 

\par Nevertheless, this formulation is not enough to characterize the spatial interactions between obstacles and transmitters. Below, we first  consider two transmitters located on the line $ y=d_1+d_2. $ 
\begin{proposition}\label{L:2}
	Suppose we have two transmitters at $ (x_1,d_1+d_2) $ and $ (x_2, d_1+d_2), $ respectively. The probability that the typical receiver is LOS w.r.t. these two transmitters is 
	\begin{equation*}
		\exp\left(-\frac{4\lambda_b}{\mu}+\frac{2\lambda_be^{-\mu(\hat{x}_2-\hat{x}_1)}}{\mu} + \lambda_b(\hat{x}_2-\hat{x}_1)e^{-\mu(\hat{x}_2-\hat{x}_1)}\right),
	\end{equation*}
where $ \hat{x}_1 {=} \frac{d_1}{d_1+d_2}x_1 $ and $ \hat{x}_2 {}{=}\frac{d_1}{d_1+d_2}x_2. $ 
\end{proposition}

\begin{IEEEproof}
	 Let $ \ind_{\text{LOS}(x_1,x_2;Y_j) } $ be the indicator that takes a value of one when none of paths from two these transmitters are blocked by any obstacle. Then, we have 
	\begin{align*}
		\bE[\ind_{\text{NLOS}(x,y)}] &= \bE[1-\ind_{\text{LOS}(x_1,x_2;y_j \forall y_j\in\Phi_b) }]\\
		&=\bE\left[1-\prod_{y_j\in\Phi_b}\bE[\ind_{\text{LOS}(x_1,x_2;y_j) }|\Phi_b]\right]\\
		&=1-\bE\left[\prod_{y_j\in\Phi_b}\bE[\ind_{\text{LOS}(x_1,x_2;y_j ) }|\Phi_b]\right].
	\end{align*}
The above conditional expression is decomposed into three intervals based on the relative locations of obstacles w.r.t. the projections of the two transmitters. See Fig. \ref{fig:picture2} for the projection. For instance, for an obstacle with $y_j< \hat{x}_1, $ its right-hand side segment length should be less than $ \hat{x}_1-y_i$ to avoid blocking the LOS path from the transmitter at $ (x_1,d_1+d_2) $ to the typical receiver.  Similarly, for an obstacle with $ \hat{x}_1<y_j<\hat{x}_2 $, its right-hand side  and left-hand side segment lengths should be less than $ \hat{x}_2-y_j $ and $ y_j-\hat{x}_2 $, respectively. Therefore, we have to evaluate the following conditional probabilities:
\begin{align*}
&\text{For all }  y_j< \hat{x}_1,  \bP\left(\tilde{W}_j<\hat{x}_1-y_j|\Phi_b\right),   \\
&\text{For all } \hat{x}_1 < y_j < \hat{x}_2,	\bP\left(\tilde{V}_j<y_j-\hat{x}_1,  \tilde{W}_j<\hat{x}_2-y_j|\Phi_b\right), \\
&	\text{For all }y_j < \hat{x}_2:  \bP\left(\tilde{V}_j<\hat{x}_2-y_j|\Phi_b\right).
\end{align*}
The three expressions can be evaluated further since $ \{V_j,W_j\} $ are i.i.d. exponential random variables with mean $ 1/\mu. $ Notice that the expression in the middle captures the condition that any obstacle whose center is within $ [\hat{x}_1, \hat{x}_2] $ should  not simultaneously block the direct paths from $ x_1 $ and $ x_2 $ at the same time.  By using the density function of the exponential distribution and by using the probability density function of the exponential random variable, the NLOS probability is given by 
\begin{align}
1-\bE_{\Phi_b}&\left[\prod_{y_j\in \Phi_b(I_1)}\left(1-e^{-\mu\left(\hat{x}_1 - y_j\right)}\right)\right.\nnb\\
	&\hspace{2.4mm}\prod_{y_j\in \Phi_b(I_2)}\left(1\!-e^{-\mu(y_j-\hat{x}_1)}\right)\left(1-e^{-\mu(\hat{x}_2-y_j)}\right)\nnb\\
	&\hspace{2.4mm}\left.\prod_{y_j\in \cap\Phi_b(I_3)}\left(1-e^{-\mu\left(y_j-\hat{x}_2 \right)}\right)\right]\nnb,
\end{align}
where 
\begin{align*}
	I_1  = \left(-\infty, \hat{x}_1\right),	I_2  = \left(\hat{x}_1, \hat{x}_2\right),	I_3  = \left(\hat{x}_2, \infty\right).
\end{align*}
Then, we use the probability generating functional of the obstacle point process to obtain the final result. 
\end{IEEEproof}

\begin{figure}
	\centering
	\includegraphics[width=1\linewidth]{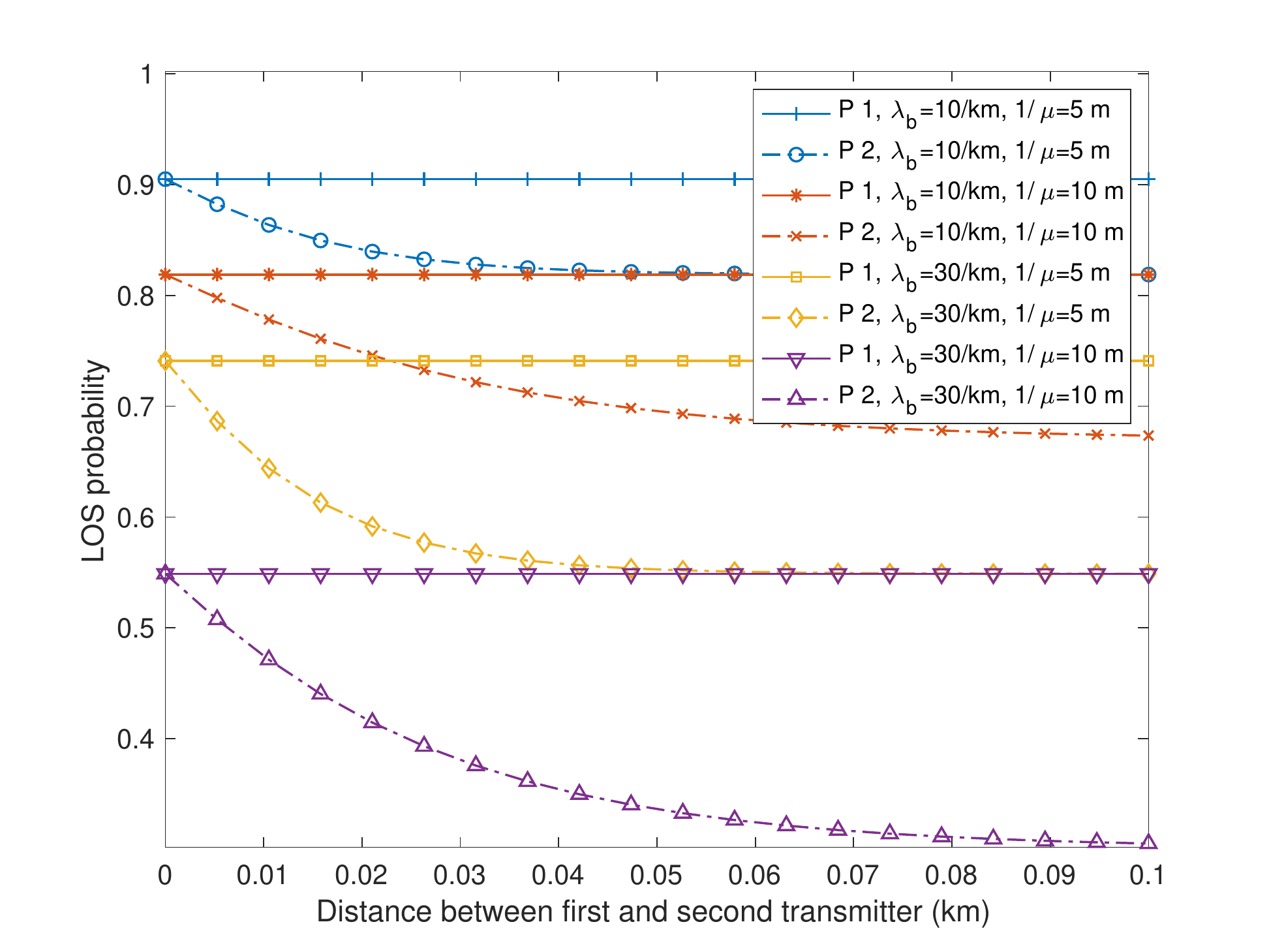}
	\caption{Illustration of derived formulas in Propositions 1 and 2.}
	\label{fig:comparisonp1p2}
\end{figure}

\begin{figure}
	\centering
	\includegraphics[width=1\linewidth]{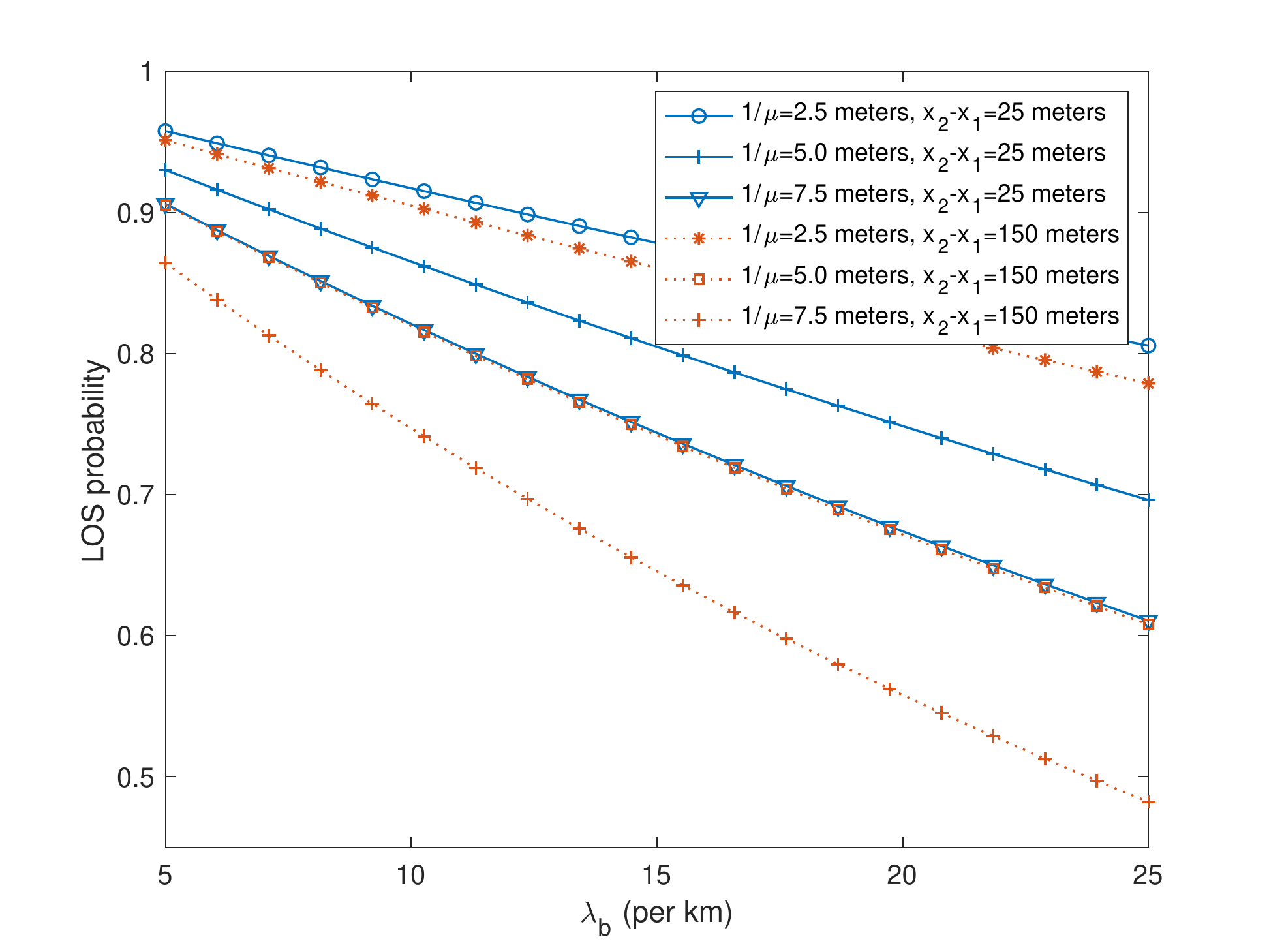}
	\caption{Illustration of the derived formula in Proposition 2 with $ d_1=10 $ meters and $ d_2=40  $ meters. }
	\label{fig:prop2}
\end{figure}

Fig. \ref{fig:comparisonp1p2} illustrates that the formulas provided in Propositions \ref{T:1} and \ref{L:2}. We consider $ d_1=d_2=10 $ meters, and the $ x $-coordinate of the first transmitters is $ 0. $ It numerically shows that when the distance $ d $ between the two transmitters---located at $ (0,20) $ and at $ (d,20) $, respectively--- is small, the probability that the typical receiver is LOS w.r.t. these two transmitters corresponds to the probability that the typical receiver is LOS w.r.t. a single transmitter. This shows that the direct paths from these two transmitters experience the same set of obstacles. Therefore, Fig. \ref{fig:comparisonp1p2} illustrates that assuming the independence across transmitter-receiver pairs is inaccurate, especially when they are close. As the distance $ d $ increases, the direct paths from two transmitters start to experience independent sets of obstacles. Therefore, the probability that the typical receiver is LOS w.r.t. two transmitters decreases. When the distance $ d $ is greater than $ 100 $ meters, it stops decreasing. In this example, the obstruction of direct paths from these two transmitters are almost independent when the distance is greater than $ 100 $ meters.
Fig. \ref{fig:prop2} plots the probability that the typical receiver is LOS w.r.t. these two transmitters as the density of obstacles varies. 

\begin{proposition}\label{P:3}
	Suppose $ n \geq 2 $ transmitters on line $ y=d_1+d_2 $ with ordered $ x $-coordinates: $ x_1< \ldots<x_n$ . The probability that the typical receiver is LOS w.r.t. all $ n $ transmitters is 
	\begin{equation}
		\exp\left(-\frac{2 \lambda_b 
		 }{\mu}+\sum_{k=2}^{n}\left ( \frac{2}{\mu}+\Delta_k\right )\lambda_be^{-\mu\Delta_k}\right),\nnb
	\end{equation}
where $ \hat{x}_k = \frac{d_1x_1}{d_1+d_2} $ for $ k=1,\ldots,n, $ and $ \Delta_k =\hat{x}_k-\hat{x}_{k-1}$ for $ k=2,\ldots,n. $
%
\end{proposition}
\begin{IEEEproof}
	The NLOS probability of the typical receiver is 
		\begin{align*}
		&\bE[\ind_{\text{NLOS}(x_1,\ldots,x_n)}] \\
		&= \bE[1-\ind_{\text{LOS}(x_1,\ldots,x_n;Y_j \forall Y_j\in\Phi_b) }]\\
		&=\bE[\bE[1-\ind_{\text{LOS}(x_1,\ldots,x_n;y_j \forall y_j\in\Phi_b) }|\Phi_b]]\\
		&=\bE\left[1-\prod_{y_j\in\Phi_b}\bE[\ind_{\text{LOS}(x_1,\ldots,x_n;y_j ) }|\Phi_b]\right]\\
		&=1-\bE\left[\prod_{y_j\in\Phi_b}\bE[\ind_{\text{LOS}(x_1,\ldots,x_n;y_j ) }|\Phi_b]\right],
	\end{align*}
where $ \bE[\ind_{\text{LOS}(x_1,\ldots,x_n;y_j ) }|\Phi_b] $ denotes the conditional expectation that the direct paths from $ x_1 , \ldots,  x_n $ are not blocked. Let $ \hat{x}_i = \frac{d_1}{d_1+d_2}x_i $ for  $i=1,\ldots, n. $ Note we have $ \hat{x}_1 < \hat{x}_2 < \ldots <\hat{x}_n $. 
\par Similar to the proof of Proposition \ref{L:2}, we write the conditional probability that obstacles do not block the direct paths as follows: 
	\begin{align*}
		&\text{For all }y_j\in \Phi_b(I_1): \bP(\tilde{W}_j < \hat{x}_1-y_j|\Phi_b) ,\\
		&\text{For all }y_j\in \Phi_b(I_2): \bP(\tilde{V}_j<y_j-\hat{x}_1, \tilde{W}_j < \hat{x}_2-y_j|\Phi_b),\\
		& \hspace{35mm}\vdots\nnb\\
		&\text{For all }y_j\in \Phi_b(I_{n+1}): \bP(\tilde{V}_j<y_j-\hat{x}_n|\Phi_b),  
	\end{align*}
	where 
	\begin{align*}
		I_1 = (-\infty, \hat{x}_1),	\ldots,	I_k = (\hat{x}_{k-1}, \hat{x}_k),\ldots,
		I_{n+1}  = (\hat{x}_n, \infty).
	\end{align*}
Similar to the proof of Proposition \ref{L:2},  we have 
\begin{align*}
	\bE_{\Phi_b}\!&\left[\prod_{y_j\in \Phi_b(I_1)}\left(1-e^{-\mu\left(\hat{x}_1-y_j\right)}\right)\right.\\
	&\left.\prod_{k=2}^{n}\left(\prod_{y_j\in \Phi_b(I_k)}\left(1-e^{-\mu(\hat{x}_{k}-y_j)}\right)\left(1-\!e^{-\mu(\hat{x}_{k-1}-y_j)}\right)\right)\right.\\
	&\left.\left.\prod_{y_j\in \Phi_b(I_{n+1})}1-e^{-\mu(y_j-\hat{x}_n)}\right.\right].
\end{align*}
We obtain the final result by using the probability generating functional of the obstacle point process of intensity $ \lambda_b $.
\end{IEEEproof}
\begin{figure*}
	\begin{align}
	&\lambda_t\xi e^{-\lambda_t\xi-2\lambda_b\mu}+\sum_{n=2}^{\infty}e^{-\lambda_t\xi}\frac{(\lambda_t\xi)^n}{n!} \underbrace{\int_{-\frac{\hat{\xi}}{2}\leq \hat{x}_1<\ldots<\hat{x}_n\leq \frac{\hat{\xi}}{2}}}_{n\text{-fold}}	 {\exp\left(-\frac{2 \lambda_b n }{\mu}+\sum_{k=2}^{n}\left ( \frac{2}{\mu}+\Delta_k\right )\lambda_be^{-\mu\Delta_k}\right)}	{\diff \hat{x}_{1}\ldots \diff \hat{x}_n}\label{main}.\\
	&	\lambda_t\xi e^{-\lambda_t\xi-2\lambda_b\mu}+ \sum_{n=2}^{\infty}e^{-\lambda_t\xi}\frac{(\lambda_t\xi)^n}{n!} \underbrace{\int_{-\frac{\hat{\xi}}{2}\leq \hat{x}_1<\ldots<\hat{x}_{n}\leq \frac{\hat{\xi}}{2}}}_{n\text{-fold}}\left(\sum_{j=1}^{n} (-1)^{j+1} \hspace{-6mm}\sum_{1\leq i_1< \ldots<i_j \leq n  }		  e^{-\frac{2 \lambda_b n }{\mu}+\sum_{m=2}^{j}\left ( \frac{2}{\mu}+\zeta_m\right )\lambda_be^{-\mu\zeta_m}}	 \right){\diff \hat{x}_{1}\ldots \diff \hat{x}_n}\label{233}.
	\end{align} 
		\vspace{-2mm}
		\\\rule{\textwidth}{0.4pt}
\end{figure*}

\subsection{LOS Coverage Probability}
We now consider the signal attenuation and detection threshold in the characterization of LOS to account for the use in wireless applications. 
Recall the typical receiver is in \emph{full LOS }coverage if it is in LOS w.r.t. all of its detectable transmitters. On the other hand, the typical receiver is in $ k $-LOS coverage if it is in LOS w.r.t. at least $ k $ detectable transmitters. 

\par First, let $ d_{\star} $ be the maximum distance at which a transmitter is detectable by the typical receiver. We have \[ d_\star = \argmax_{d}\left\{\frac{p d^{-\alpha_{\text{LOS}}}}{\sigma}>\tau \right\}= \left(\frac{p}{\sigma\tau}\right)^{\frac{1}{\alpha_{\text{LOS}}}}, \]
where $ \tau $ is the minimum SNR detection threshold at receivers. 

\begin{theorem}The full LOS coverage probability of the typical receiver is given by Eq. \eqref{main}. 
\end{theorem}
\begin{IEEEproof}
	Let $ \Xi $ denote the intersection of a ball of radius $ d_\star $ centered at the origin and the line $ y=d_1+d_2. $ Then, the LOS coverage probability is given by 
	\begin{align}
\sum_{n=1}^{\infty}\bP(\text{full LOS cov}|\Phi_r(\Xi)=n )\bP(\Phi_r(\Xi)=n),\label{2}
	\end{align}
where $ \bP(\text{full LOS}|\Phi_r(\Xi)=n) $ is full LOS coverage conditional on  $ n $ detectable transmitters within a distance $ d_\star $ from the origin. We have  
 \begin{align*}
	\bP(\Phi_r(\Xi)=n) = e^{-\lambda_t\xi}\frac{\left(\lambda_t\xi\right)^{n}}{n!},
\end{align*}
where $ \xi = 2\sqrt{d_\star^2-(d_1+d_2)^2}. $ 
\par On the other hand, the conditional expression in Eq. \eqref{2} is 
	\begin{align}
	&\bE[\ind_{\text{LOS}(X_1,\ldots,X_n)}|n]\nnb\\
	&=\bE_{X_{[n]}}\left[\bE_{\Phi_b}\left[\prod_{y_j\in\Phi_b}\bE[\ind_{\text{LOS}(x_{[n]};y_j ) }|\Phi_b,{X_{[n]}},n]\right]\right]\nnb\\
	&=\bE_{X_{[n]}}\left[\bE_{\Phi_b}\left[\prod_{y_j\in\Phi_b}\bE[\ind_{\text{LOS}(x_{[n]};y_j) }|\Phi_b,{X_{[n]}},n]\right]\right],\label{eq:4}
\end{align}
where $ \ind_{\text{LOS}(x_1,\ldots,x_n; y_j) } $ is $ 1 $ if none of the direct paths from the transmitters $ x_1 $ to $ x_n $ are blocked by the obstacle centered at $ y_j $. We denote by $ X_{[n]} $ the locations of the $ n $ transmitters. 
To have \eqref{eq:4}, we condition on the the locations of the $ n $ transmitters and the locations of obstacles. 
Note the outermost expectation is w.r.t. the joint distribution of the ordered locations of the $ n $ points on the segment of length $ \xi $: $ X_{[n]}\equiv\{{X}_{1},\ldots,{X}_n\} $. For their  joint distributions, see \cite[pp. 24]{daley2003introduction}

Based on the conditioning on  $ n $ transmitters, the innermost expectation coincides with the LOS probability derived in Proposition \ref{P:3}. Therefore,  the conditional expression of Eq. \eqref{2}  is given by 
\begin{align}
			\bE_{X_{[n]}}\left[		\exp\left(-\frac{2 \lambda_b n }{\mu}+\sum_{k=2}^{n}\left ( \frac{2}{\mu}+\Delta_k\right )\lambda_be^{-\mu\Delta_k}\right)\right],\nnb
\end{align}
where the expectation is w.r.t. the joint distribution of the locations of $ n $ projections, namely $ \hat{x}_{[n]}\equiv\{\hat{x}_1,\ldots,\hat{x}_n \text{ where }  \hat{x}_i = \frac{d_1}{d_1+d_2}x_i \} $.
\end{IEEEproof}

\begin{figure}
	\centering
	\includegraphics[width=1\linewidth]{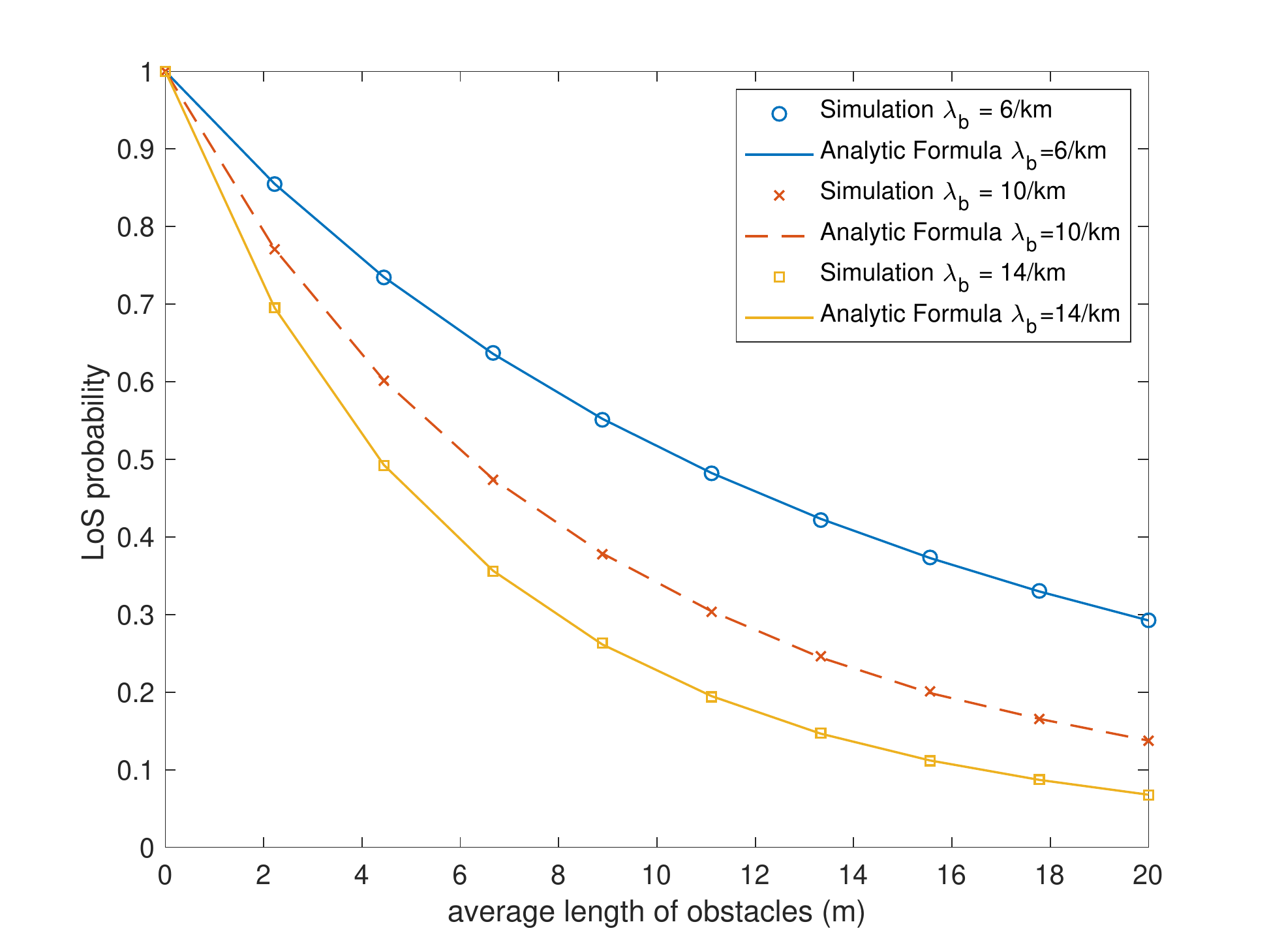}
	\includegraphics[width=1\linewidth]{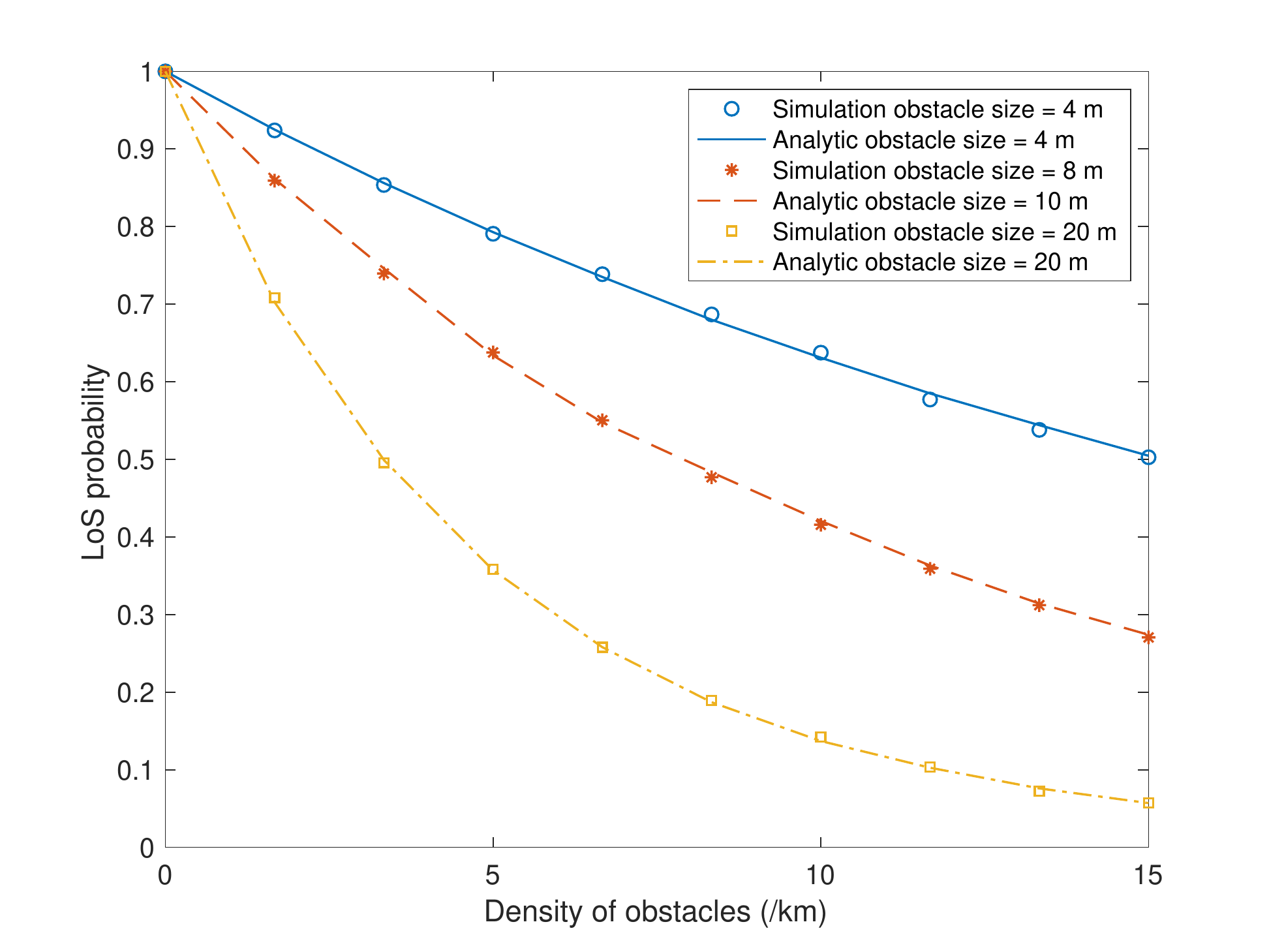}
	\caption{Illustration of the full LOS coverage probability. We consider $ d_1=10 $ m, $ d_2=10 $ m, and $ d_\star=1.5 $ km. We use $ \lambda_t=4/\text{km}$ to have the average inter-transmitter distance as $ 250 $  meters.}
	\label{fig:losprobability}
\end{figure}

The LOS probabilities for various densities of obstacles and transmitters are provided in Fig. \ref{fig:losprobability}. It shows that the derived formula matches the simulation results for various parameters.  We use $ \lambda_b = \{6, 10, 14\} $ per kilometer, and these values correspond to the inter-obstacle distances of $ 166, 100,$ and $71 $ meters on average, respectively. As the average length of obstacles increases, the LOS coverage probability decreases. 
\begin{theorem}\label{T:2}
The probability that the typical receiver is LOS w.r.t. at least one transmitter is given by Eq. \eqref{233}. 
\end{theorem}
\begin{IEEEproof}
Conditional on the presence of $  n $ transmitters on the segment $ \Xi $ and conditional on their ordered $ x $-coordinates, denoted by $ x_1<\cdots<x_n $, the $ 1 $-LOS coverage probability is 
\begin{align*}
	\bP(1\text{-LOS})= \sum_{n=0}^{\infty}&\bP(1\text{-LOS}|\Phi_r(\Xi)=n )\bP(\Phi_r(\Xi)=n)\\
	=\sum_{n=1}^{\infty}&\bP\left(\left.\bigcup_{i=1}^{n}L_i\right|\Phi_r(\Xi)=n \right)\bP(\Phi_r(\Xi)=n)\\
	=\sum_{n=1}^{\infty}&\bE\left[\bP\left(\left.\bigcup_{i=1}^{n}L_i\right| X_{[n]},\Phi_r(\Xi)=n \right)\right]\\
	&\times \bP(\Phi_r(\Xi)=n).
\end{align*}
	Here, $ L_i $ denotes the event that the typical receiver is LOS w.r.t. the transmitter indexed by $ i $. Then, based on the inclusion-exclusion formula \cite{van2001course}, we have 
\begin{align}
	&\bP\left(\left.\bigcup_{i=1}^{n} L_i \right| X_{[n]}, \Phi_r(\Xi)=n\right)\nnb\\
	&= \sum_{j=1}^{n}(-1)^{j+1} \hspace{-8mm}\sum_{\{1\leq i_1< i_2<\ldots<i_j \leq n  \}}\hspace{-8mm}\bP\left(L_{i_1,i_2,\ldots,i_j}|X_{[n]},\Phi_r(\Xi)=n\right),\nnb
\end{align}
where $ L_{i_1,i_2,\ldots,i_j} $ is the event that the typical receiver at the origin is LOS w.r.t. the transmitters $ i_1, \ldots, i_j$.

As in proof of Proposition \ref{P:3}, conditional on $ n $ transmitters and their ordered locations, the typical receiver is LOS w.r.t. all the transmitters $ i_i, \ldots,i_j $ iff 
 \begin{align*}
 	&\forall y_k\in \Phi_b(I_{i_1}): \tilde{W}_k<\hat{x}_{i_1}-y_k,\\
 	&\forall y_k\in \Phi_b(I_{i_2}): \tilde{W}_k<y_k - \hat{x}_{i_1}, \tilde{V}_k < \hat{x}_{i_2}- y_k,\\
 	&\hspace{5mm}\vdots \\ 
 	&\forall y_k\in \Phi_b(I_{i_{j+1}}): \tilde{V}_k<y_k-\hat{x}_{i_j},
 \end{align*}
where 
\begin{equation*}
	I_{i_1} = (-\infty, \hat{x}_{i_1}), I_{i_2} = (\hat{x}_{i_1} , \hat{x}_{i_2}), \ldots, I_{i_{j+1}} = (\hat{x}_{i_j},\infty ),
\end{equation*}
and $\hat{x}_{i_j}$ is the $ x $-coordinate of the projection of the transmitter at $ (x_{i_j},d_1 + d_2) $. Then, $ \bP\left(L_{i_1,i_2,\ldots,i_j}|X_{[n]},\Phi_r(\Xi)=n\right) $ is 
\begin{align*}
		&\bE_{\Phi_b}\!\left[\prod_{y_k\in \Phi_b(I_{i_1})}\left(1-e^{-\mu\left(\hat{x}_{i_1}-y_k\right)}\right)\right.\\
		&\hspace{9mm}\left.\prod_{m=2}^{j}\!\left(\prod_{y_k\in \Phi_b(I_{i_m})}\left(\!1-e^{-\mu(\hat{x}_{i_m}-y_k)}\right)\left(\!1-\!e^{-\mu(\hat{x}_{i_{m-1}}-y_k)}\right)\right)\right.\\
		&\hspace{6mm}\left.\left.\prod_{y_k\in \Phi_b(I_{i_{j+1}})}\left(1-e^{-\mu(y_k-\hat{x}_{i_j})}\right)\right.\right]\\
		&=\exp\left(-\frac{2 \lambda_b 
		}{\mu}+\sum_{m=2}^{j}\left ( \frac{2}{\mu}+\zeta_m\right )\lambda_be^{-\mu\Delta_m}\right),
	\end{align*}
where $ \zeta_m = \hat{x}_{i_m}-\hat{x}_{{i_{m-1}}} $ for $ m=2,\ldots, j. $  We get the final result by deconditioning w.r.t. the ordered locations of $ n $ transmitters, and w.r.t. the number of transmitters.
\end{IEEEproof}
\begin{example}
The probability that the typical receiver is LOS w.r.t. at least two transmitters can be derived as in Theorems \ref{T:1} and \ref{T:2}. Let $ L_{i,j} $ denote the event that the typical receiver is in LOS coverage w.r.t. the two transmitters $ i  $ and $ j. $ Conditionally on $ n $ transmitters on the segment $ \Xi $ and their ordered $ x $-coordinates, namely $ x_1<\cdots<x_n $, the event that the typical user is LOS w.r.t. at least two transmitters is given by $ \bigcup\limits_{1\leq i< j\leq n}L_{i,j}. $ From the inclusion-exclusion formula, 
	\begin{align}
		&\bP\left(\bigcup\limits_{1\leq i< j\leq n}L_{i,j}| X_{[n]}, \Phi_r(\Xi)=n \right)\nnb\\
		& = \sum_{m=1}^{\frac{n(n-1)}{2}} (-1)^{m+1}  \sum_{\cS}\bP\left(L_{i_1,j_1,\ldots,i_m,j_m}|X_{[n]}, \Phi_r(\Xi)=n\right),\label{eq:5}
	\end{align}
where $  \cS= 1,2\leq i_1,j_1 < \ldots < i_m,j_m\leq n-1,n   $ in the lexicographical order and $ L_{i_1,j_1,\ldots,i_m,j_m} $ is the event that the typical receiver is LOS w.r.t. the transmitters $ i_1,j_1,\ldots,i_m,j_m$. Suppose transmitters indexed in Eq. \eqref{eq:5} corresponds to $ k $ distinct transmitters.  Then, using their projections, one can derive the set of conditions that the typical receiver is in LOS w.r.t the $ k $ transmitters as in Theorems \ref{T:1} and \ref{T:2}.
\par
Finally, one have the final result by (i) exploiting the probability density function of the exponential random variable, (ii) using the probability generating functional of the obstacle point process, (iii) deconditioning w.r.t. the ordered locations of the $ n $ transmitters, and (iv) deconditioning w.r.t. the number of transmitters in the segment of length $ \xi. $  
\end{example}

\section{Conclusion}
This paper proposes a spatially consistent random geometric model to identify blockage and LOS in vehicular networks. This paper uses a simple stochastic geometry model for vehicular networks to characterize the spatially correlated LOS paths in the presence of obstacles. Quantifying the geometric interactions between transmitters, receivers, and obstacles, we derive the probability that a typical receiver is in the LOS. Then we evaluate the LOS coverage probability to account for the signal attenuation and the detection threshold in practice. This paper will be useful to the accurate evaluation of LOS-critical applications in vehicular networks such as positioning of vehicles or mmWave communications. 
%
%

%
\section*{Acknowledgment}
The work of C.-S. Choi was supported in part by the NRF-2021R1F1A1059666 and by the Hongik University New Faculty Research Fund. The work of  F. Baccelli was supported in part by the Simons Foundation grant \#197982 and by the ERC NEMO grant \#788851 to INRIA.

\bibliographystyle{IEEEtran}
\bibliography{ref}
\end{document}